\documentclass[10pt,letterpaper]{article}
\usepackage{opex3}

\begin{document}

\title{Super-critical phasematching for photon pair generation in structured light modes}

\author{Rebecca Y. Saaltink,$^{1,*}$ Lambert Giner,$^{1}$ Robert W. Boyd,$^{1,2}$ Ebrahim Karimi,$^{1}$ and Jeff S. Lundeen$^1$}

\address{$^1$ Department of Physics, University of Ottawa, 25 Templeton Street, Ottawa, Ontario K1N 6N5, Canada\\
$^2$ The Institute of Optics and Department of Physics and Astronomy, University of Rochester, Rochester NY 14627, USA\\}

\email{$^*$ rsaal066@uottawa.ca} 



\begin{abstract}
We propose a method for directly producing radially and azimuthally polarized photon pairs through spontaneous parametric downconversion (SPDC). This method constitutes a novel geometry for SPDC, in which a radially polarized Bessel-Gauss pump beam is directed into a nonlinear crystal, with the central propagation direction parallel to the crystal axis. The phasematching conditions are controlled by changing the opening angle of the pump beam; as the crystal axis cannot be tuned, we refer to this process as \textit{super-critical phasematching}. We model and plot the spatial and polarization output distributions for Type-I and Type-II super-critical phasematching.
\end{abstract}



\section{Introduction}
Beams displaying radial and azimuthal polarizations have drawn great interest for their unique properties and uses in applied and fundamental optics. Radially polarized light beams have polarizations aligned radially toward the beam propagation axis. They produce a strong longitudinal electric field with a focal spot below the diffraction limit under tight focusing \cite{quabis2000focusing,dorn2003sharper}. Azimuthally polarized beams have polarizations at all points orthogonal to the beam radius. When focused, they  produce a strong longitudinal magnetic field \cite{wang2008creation}. These two polarization modes are mutually orthogonal and display cylindrical symmetry about the beam axis. Radially and azimuthally polarized photons have gained increased interest in the quantum regime for applications in quantum information, such as alignment-free QKD \cite{vallone2014free,d2012complete}, single photon spin-orbit non-separable states \cite{karimi2010spin}, superdense coding and quantum communication \cite{Milione:13}, as well as in the classical regime in optical data storage and optical lithography. When used along with standard TEM$_{00}$ modes in mode-division multiplexing they have been shown to increase the bandwidth of telecommunications channels \cite{Bozinovic28062013, Milione:15}.

For propagation in free space, radially and azimuthally polarized Bessel-Gauss beams are solutions to the vectorial form of the paraxial wave equation \cite{Jordan:94}. These beams have a transverse profile given by the Bessel function of the first kind, J$_1$, multiplied by a Gaussian factor, since ideal Bessel beams are not physically achievable. They are desirable in optical applications because they exhibit non-diffracting propagation  with an improved depth of focus over Gaussian beams \cite{mcgloin2005bessel, greene1996diffraction}. It is key to our modeling that a Bessel-Gauss beam can be decomposed into a distribution of Gaussian spatial mode beams with their central wave-vectors along the surface of a cone \cite{schimpf2013radially}. 

We are interested in producing radially and azimuthally polarized photon pairs through spontaneous parametric downconversion (SPDC). This is an optical process that occurs within a $\chi^{(2)}$ nonlinear crystal, in which one pump photon spontaneously decays to produce two photons at lower frequencies. These two photons, commonly referred to as signal and idler, are correlated in position and momentum, and time and energy due to momentum-energy conservation \cite{walborn2010spatial}. Conservation of energy dictates that the frequencies of the two downconverted photons must add up to the pump frequency:

\begin{equation}
\omega_p = \omega_s + \omega_i.
\label{Eq:EnergyConservation}
\end{equation}

\noindent
Here, $\omega_p$, $\omega_s$, and $\omega_i$ are the frequencies of the pump, signal, and idler photons, respectively. In this paper we will restrict our attention to a CW pump laser, i.e. in effect, a single frequency. Downconverted photons may be produced with some $\mathbf{k}$-vector mismatch $\Delta \mathbf{k}$:

\begin{equation}
\mathbf{k}_s + \mathbf{k}_i - \mathbf{k}_p = \Delta \mathbf{k}.
\label{Eq:MomentumConservation2}
\end{equation}

\noindent
Here, $\mathbf{k}_p$, $\mathbf{k}_s$, and $\mathbf{k}_i$ are any set of pump, signal, and idler $\mathbf{k}$-vectors. When $\Delta \mathbf{k}$ is zero, we have perfect phasematching, i.e. momentum conservation.			

In this paper, we focus on Type-II phasematching (a discussion on Type-I phasematching can be found in Appendix A1). In Type-II phasematching, the signal and idler polarizations are orthogonal. The downconverted photons may be emitted in different directions or, in the case of collinear downconversion, in the same direction. The wavelengths and emission directions of these photons depend on the angle of the pump beam relative to the optic axis of the crystal, and are constrained by the conservation rules given above in Eqs. (\ref{Eq:EnergyConservation}) and (\ref{Eq:MomentumConservation2}). In the opposite process, sum-frequency generation, the amount of light produced is critically sensitive to the angle of the crystal. Hence, this type of phasematching is called \textit{critical phasematching}.

We propose a novel geometry for SPDC that will produce photon pairs with one radially polarized and one azimuthally polarized photon. The pump is a radially polarized Bessel-Gauss beam that is focused into the crystal such that its cone axis is parallel to the crystal optic axis.  The opening angle of the pump cone is chosen such that the central \textbf{k}-vector of each Gaussian beam in the pump distribution satisfies the phasematching conditions in Eq. (\ref{Eq:MomentumConservation2}). In this geometry, phasematching is achieved by adjusting the pump cone angle or, if possible, the crystal temperature. Since even the crystal angle cannot be adjusted anymore, we refer to this process as \textit{super-critical phasematching}. In this paper, we model the spatial and polarization properties of the generated photon pairs.

\section{Geometry and notation}

\begin{figure}[h!t!p!b!]
\centering
  \makebox[\textwidth][c]{\includegraphics[width=1\textwidth]{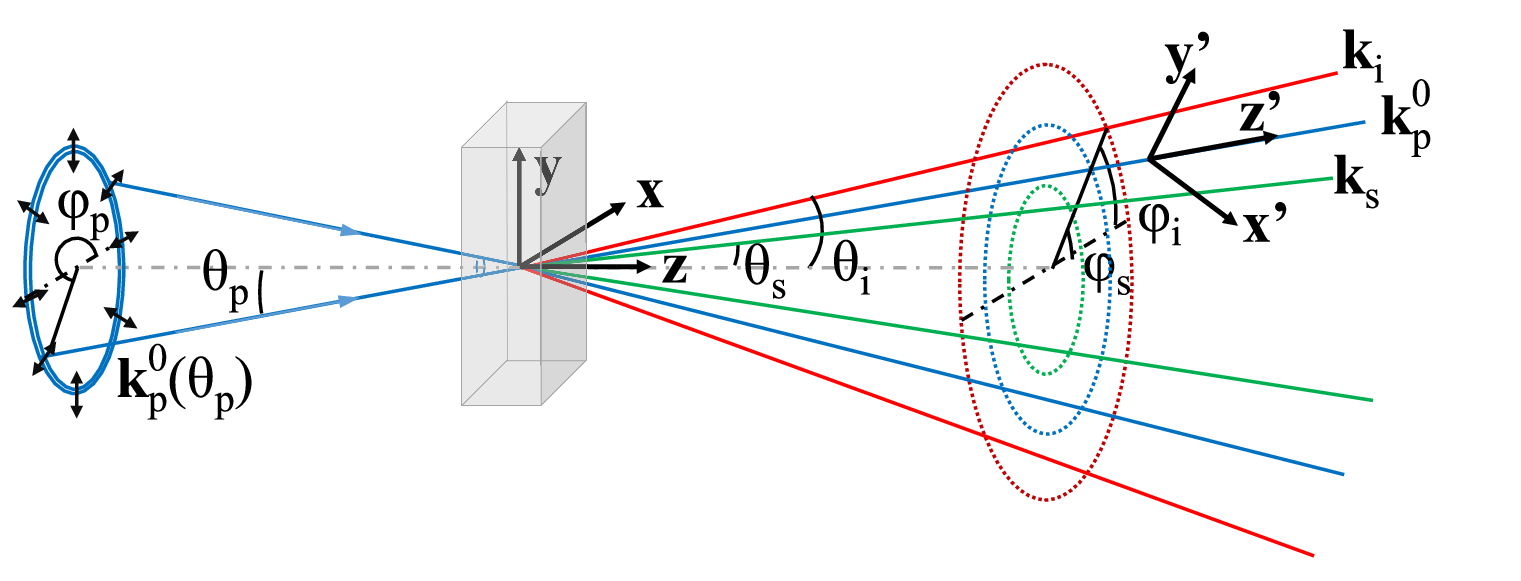}}
\caption{The geometry of super-critical phasematching. The $\mathbf{x}$, $\mathbf{y}$, and $\mathbf{z}$ axes are the crystallographic axes, where $\mathbf{z}$ is the optic axis. The pump (blue) is comprised of a conical distribution of Gaussian beams, each with central $\mathbf{k}$-vector $\mathbf{k}_p^0(\varphi_p)$, where $\varphi_p$ is the azimuthal angle. The pump beam cone axis is along the crystal axis $\mathbf{z}$, and the opening half-angle of the cone is $\theta_p$. The signal (idler), shown in green (red), is emitted with $\mathbf{k}$-vector $\mathbf{k}_{s(i)}$. The signal (idler) $\mathbf{k}$-vector is characterized by two angles: $\theta_{s(i)}$, the half-cone opening angle, measured between the signal (idler) $\mathbf{k}$-vector and the $\mathbf{z}$ axis; and $\varphi_{s(i)}$, the azimuthal angle. We also introduce a local frame, denoted by $\mathbf{x}$', $\mathbf{y}$', and $\mathbf{z}$', where $\mathbf{z}$' is defined along the central $\mathbf{k}$-vector, $\mathbf{k}_p^0$, of each Gaussian beam. The other axes are aligned azimuthally ($\mathbf{x}$') and radially ($\mathbf{y}$') with respect to $\mathbf{z}$.}
\label{fig:SPDCGeo}
\end{figure}

In this section, we fully describe the proposed geometry for super-critical phasematching. We define the crystallographic axes by $\mathbf{x}$, $\mathbf{y}$, and $\mathbf{z}$,  as shown in Fig. \ref{fig:SPDCGeo}, where $\mathbf{z}$ is the optic axis (we restrict our analysis to a uniaxial crystal). We consider degenerate phasematching, i.e. the signal and idler wavelengths are the same. 

As mentioned in the introduction, a Bessel-Gauss pump beam, BG($\mathbf{k}$), can be thought of as a distribution of  Gaussian beams around a cone \cite{schimpf2013radially}:

\begin{equation}
BG(\mathbf{k}) \propto \int_{0}^{2\pi} e^{-w_{p}^{2}(\mathbf{k}-\mathbf{k}_p^{0}(\varphi_p))^{2}/4}d\varphi_p,
\label{Eq:Bessel}
\end{equation}
\begin{equation}
k^{0}_{p,x}(\varphi_p)=|\mathbf{k}_p^{0}|\sin(\theta_p)\cos(\varphi_p), \ \ \ \  k^{0}_{p,y}(\varphi_p)=|\mathbf{k}_p^{0}|\sin(\theta_p)\sin(\varphi_p),\ \ \ \ 
k^0_{p,z} = |\mathbf{k}_p^{0}|\cos(\theta_p)
\label{Eq:Cone}
\end{equation}

\noindent
where each of these Gaussian beams has a central \textbf{k}-vector given by $\mathbf{k}_p^0(\varphi_p)$ (with components $k^{0}_{p,j}$ for $j=x,y,z$) ,  $w_p$ is the $1/e^2$ spatial full-width of the pump beam, $\theta_{p}$  is the opening half-angle of the cone, and $\varphi_{s(i)}$ is the azimuthal angle. Thus, the cone axis is parallel to the $z$-axis the crystal. Decomposing the Bessel-Gauss beam into this distribution has the advantage that each Gaussian beam may be treated in the paraxial approximation, whereas the full Bessel-Gauss beam can be non-paraxial if the cone angle is large. We use a proportionality symbol in Eq. (\ref{Eq:Bessel}) and later in the paper since we will impose an overall normalization later.

The \textbf{k}-vectors of the signal and idler photons are denoted by $\mathbf{k}_s$ and $\mathbf{k}_i$, respectively. The emission directions of these photons are characterized by two angles, $\theta_{s(i)}$ and $\varphi_{s(i)}$, of spherical coordinates. The first, $\theta_{s(i)}$, is measured between the \textbf{k}-vector of the signal(idler) photon and the $z$-axis of the crystal. The azimuthal angle, $\varphi_{s(i)}$, represents a rotation from the $x$-axis in the transverse plane of the crystal. The angles of the pump, signal, and idler photons are defined in the crystal frame of reference, that is, with respect to $\mathbf{x}$, $\mathbf{y}$, and $\mathbf{z}$. These angles are shown in Fig. \ref{fig:SPDCGeo}.

We also define a rotated local frame given by $\mathbf{x}'$, $\mathbf{y}'$ and $\mathbf{z}'$. Here, local means that the axes are defined with respect to the central pump $\mathbf{k}$-vector $\mathbf{k}_p^0$ of a Gaussian beam in our distribution in Eq. (\ref{Eq:Bessel}), such that $\mathbf{z}'$ is in the same direction as $\mathbf{k}_p^0$. The $\mathbf{y}'$ axis is in the plane of $\mathbf{z}$ and $\mathbf{z}$'. The $\mathbf{x'}$ axis is perpendicular to this plane. Consequently, the local frame rotates in the integral in Eq. (\ref{Eq:Bessel}). That is, for every value of the integrand the axes are aligned azimuthally ($\mathbf{x'}$) and radially ($\mathbf{y'}$) with respect to $\mathbf{z}$.  Conversely, the axes $\mathbf{x}$, $\mathbf{y}$, $\mathbf{z}$ are stationary with respect to the crystal, as they are defined with respect to the optic axis of the crystal. The coordinate transformations between $\mathbf{x}$, $\mathbf{y}$, $\mathbf{z}$ and $\mathbf{x}'$, $\mathbf{y}'$, $\mathbf{z}'$ are given in Eq. (\ref{Eq:crystaltopumpframe}) from Appendix A2, and are from Ref. \cite{boeuf2000calculating}. 

As an example, we consider a negative uniaxial crystal (e.g. $\beta$-barium borate (BBO)) in which the only phasematched process occurs for an extraordinarily (e) polarized pump beam. Extraordinary polarization is defined to be in the plane formed by a beam's wavevector $\mathbf{k}$ and the crystal optic axis $\mathbf{z}$, whereas ordinary (o) polarization is orthogonal to this plane. In terms of our local frame, the pump polarization is along $\mathbf{y'}$. As $\mathbf{y'}$ is aligned radially with respect to $\mathbf{z}$, the extraordinary polarized Gaussian beams in the pump distribution in Eq.  (\ref{Eq:Bessel}) form a Bessel-Gauss beam that is radially polarized.  

\section{Simulation results}

\begin{center}
\begin{figure}[h]
\makebox[\textwidth][c]{\includegraphics[width=1\textwidth]{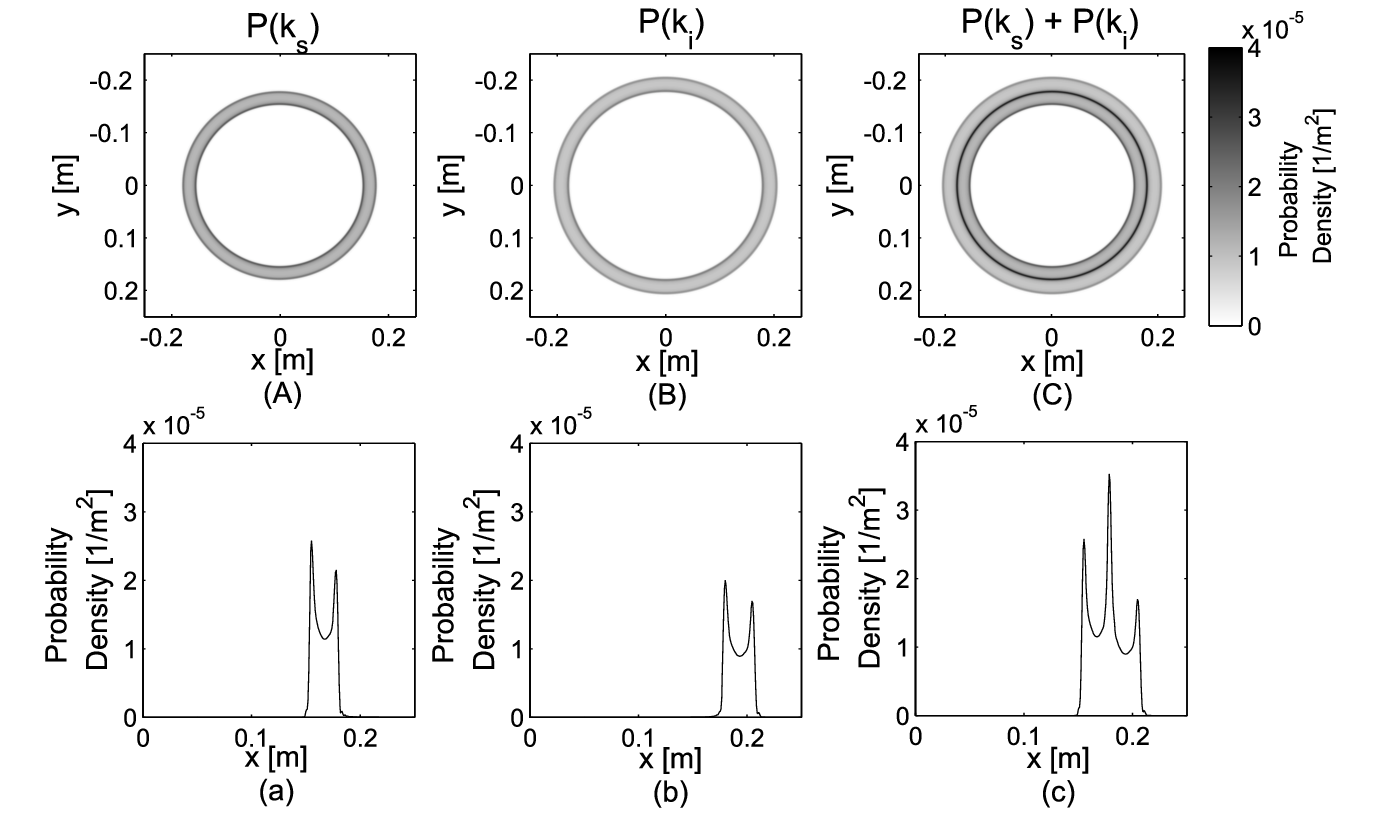}}
\caption{Marginal probability density ($P(\textbf{k}_s)$, $P(\textbf{k}_i)$) for Type-II degenerate down-conversion. The pump cone half-angle is $41.8^\circ$ in a $500\ \mu \textrm{m}$-thick BBO crystal. The pump beam wavelength is 405 nm and the signal and idler wavelengths are 810 nm. Since for degenerate phasematching $|\mathbf{k}_{s(i)}|$ is fixed, $\theta_{s(i)}$ and $\phi_{s(i)}$ completely determine the signal and idler wavevectors. In turn, the position of the photons after the crystal determines these angles, so these plots are given in the crystal frame (x,y) at z=20cm after the crystal exit face. Plots (A), (B), and (C) give (A) the probability density of generating a signal photon at $\mathbf{k}_s$, (B) the probability density of generating an idler photon at $\mathbf{k}_i$, and (C) $P(\mathbf{k}_s) + P(\mathbf{k}_i)$, which is proportional to the photon flux. Plots (a), (b), and (c) are produced from plots (A), (B), and (C), respectively, by plotting along $y=0$.}
\label{fig:TypeIIOutput}
\end{figure}
\end{center}

We give a brief description of our methods here; a full description of our calculations can be found in Appendix A2. In this section, we simulate the output \textbf{k}-vector distributions of the downconverted photons for a super-critically phasematched Bessel-Gauss pump beam.  This type of beam can be strongly non-paraxial, which means standard methods for modeling the output distribution can be difficult to apply. Instead, we begin by modelling the output from a single Gaussian pump centered on  $\mathbf{k}_p^0(\varphi_p)$. That is, we model the complex amplitude $\Phi_{G}(\mathbf{k}_p^0(\varphi_p), \mathbf{k}_s, \mathbf{k}_i)$ for the signal and idler photons to have wavevectors $\mathbf{k}_s$ and $\mathbf{k}_i$. Then we superpose these amplitudes similar to in Eq. (\ref{Eq:Bessel}) to find the total complex amplitude output from a Bessel-Gauss pump:

\begin{equation}
\Phi_{BG}(\mathbf{k}_s, \mathbf{k}_i) = \frac{1}{\sqrt{N}}\int_{0}^{2\pi} \Phi_{G}(\mathbf{k}_p^0(\varphi_p), \mathbf{k}_s, \mathbf{k}_i) d\varphi_p,
\label{Eq:PhiBG}
\end{equation}

The scaling,$\sqrt{N}$ ensures that the joint probability distribution $|\Phi_{BG}|^2$  is normalized. In \cite{boeuf2000calculating}, Boeuf et al. demonstrate that the complex amplitude $\Phi_{G}$ can be expressed as : 

\begin{equation}
\Phi_G(\mathbf{k}_p^0(\varphi_p), \mathbf{k}_s, \mathbf{k}_i) \propto \exp\left({\frac{-w_{p}^{2}(\Delta k_{x'}^2 + \Delta k_{y'}^2)}{4}}\right)\textrm{sinc}\left(\frac{L_{\mathrm{optic}}\Delta k_{z'}}{2}\right)\exp\left({\frac{iL_{\textrm{optic}}\Delta k_{z'}}{2}}\right),
\label{Eq:PhiG}
\end{equation}

\noindent
in which $\Delta k_{x'}$, $\Delta k_{y'}$, and $\Delta k_{z'}$ (given in Appendix A2) are the \textbf{k}-vector mismatches in the $\mathbf{x}'$, $\mathbf{y}'$, and $\mathbf{z}'$ directions. Here, $L_{\mathrm{optic}}$ is the length of the optical path through the crystal. That is, $L_{\mathrm{optic}}=L_{\mathrm{crystal}}/\mathrm{cos}(\theta_p)$, where $L_{\mathrm{crystal}}$ is the crystal length as measured along its normal. The first exponential term and sinc term give the square root of the phasematching function from \cite{boeuf2000calculating}, and the second exponential term gives the phase. 

The pump cone half-angle $\theta_p$ is set to $41.8^\circ$ inside a $500\ \mu m$ thick BBO crystal. This crystal thickness ensures that the sinc width in Eq. (\ref{Eq:PhiG}) is sufficiently large to capture in a numerical simulation, and the angle, $\theta_p$, is the angle for Type-II degenerate collinear phasematching at 405 nm for a typical SPDC geometry. This value of the pump angle corresponds to a typical angle used to produce photons within the sensitivity range of silicon detectors while still having a pump at a typical laser wavelength. More specifically, with a pump wavelength of 405 nm and in the case of degenerate SPDC, the signal and idler wavelengths are 810 nm. All of these parameters are typical in an SPDC experiment. Thus, every Gaussian beam in the pump distribution satisfies phasematching conditions within the BBO crystal. Each Gaussian pump beam has a $1/e^2$ spatial full-width of 84 $\mu$m. We have made the assumption that the pump, signal, and idler photons are each at a single wavelength, which is reasonable if the pump beam has a sufficiently narrow bandwidth. 

Rather than plotting the joint probability density $|\Phi_{BG}(\mathbf{k}_s, \mathbf{k}_i)|^2$   for signal and idler, which is made difficult by computation power limitations, we plot the marginal probability distribution for the signal \textit{or} idler photon. For example, we plot $P(\mathbf{k}_s) = \int\int|\Phi_{BG}|^2dk_{i,x}dk_{i,y}$ . The results are shown in Fig. \ref{fig:TypeIIOutput}. The resulting output distribution shows three concentric circles which we call ``supercones" (for reasons that will be apparent later). Signal photons are produced along the innermost and central supercones, and idler photons are produced along the central and outermost supercones. 

Implicit in our calculations are the polarization properties of the signal and idler photons. In Type-II phasematching, one downconverted photon (signal) is extraordinarily polarized and the other photon (idler) is ordinarily polarized. Thus for each $\mathbf{k}_p^0(\varphi_p)$ in Eq. (\ref{Eq:PhiBG}) the signal is polarized along $\mathbf{y'}$ and the idler is polarized along $\mathbf{x'}$. It follows that the signal output supercones are polarized radially and the idler output supercones are polarized azimuthally with respect to the $\mathbf{z}$-axis. However, there is an effect that may degrade the signal and idler's polarization. In the supercone each point arises from a small but finite range of pump angles, $\Delta \phi_p$.  Taking into account the Type II cone thickness, we quantitatively estimate $\Delta \phi_p$. For example, for a point in the overlapping signal and idler supercones $\Delta \phi_p$ will vary by $0.8^\circ$. This will also be the variation in polarization angle around the nominal polarization (e.g. radial) of the generated photons. This small effect will be further diminished since Eq. (\ref{Eq:PhiBG}) sums over that variation. We conclude that, to an excellent approximation, the generated photons have radial and azimuthal polarizations.

At this point, we do not include the effect of refraction at the input or output crystal face. If the the input and and output faces are perpendicular to the $\mathbf{z}$-axis then Snell's law would imply that the opening angles $\theta$ in free space outside the crystal will be greater than those inside. The overall shape of the distributions shown in Fig. \ref{fig:TypeIIOutput} would be retained, but refraction would increase the radii of the rings. This refraction will need to be taken into account for the experimental realization of this work. Because the phasematching angle is chosen to produce a signal and idler collinear with the pump, the large incidence angle at the crystal output face would cause the signal and idler to be separated in angle by $2.9^\circ$, which would cause them to spatially separate as they propagate. However, this is simple to precompensate by changing the phasematching angle so that they are generated inside the crystal with a  $2.9^\circ$ angular separation. Propagating across the crystal, this itself will cause a spatial walkoff. But, this effect is small compared to the pump focus width and can, thus, be neglected. These concerns are addressed further in Section 6.

In summary, by pumping with a Bessel-Gauss beam parallel to the crystal's optical axis one can generate radially and azimuthally polarized photon pairs in cylindrically symmetric spatial modes. It is important to note, these supercones are not the cone-like output distributions normally associated with critically phasematched SPDC. Namely, the supercones are cylindrically symmetric about the crystal's optical axis  $\mathbf{z}$, whereas the latter are centered on the pump axis  $\mathbf{z'}$. However, to understand how these supercones come about, in the next section we start by considering the typical output distributions for critically phasematched SPDC.

\section{Expected output}

In this section, we give a geometric argument for the spatial distributions for the signal and idler photons shown in Fig. \ref{fig:TypeIIOutput}. We do this by considering Type-II critical phasematching of a single Gaussian beam in the pump distribution in Eq. (\ref{Eq:Bessel}). By considering only a single Gaussian pump we revert to the standard geometry for SPDC, in which the pump beam travels at an angle $\theta_p$ to the crystal's optical axis.

The phasematching conditions for this Gaussian pump beam may be met by a range of signal and idler emission angles. In the simulation, we considered the case where  $\theta_p$ is set to $41.8^\circ$. For degenerate SPDC with a 405 nm pump this produces an idler and signal output cone at 810 nm, each meeting tangentially at the pump $\mathbf{k}$-vector, $\mathbf{k}_p^0$ . We call these the standard Type-II cones.

\begin{figure}[h!t!b!p!]
\centering
\makebox[\textwidth][c]{\includegraphics[width=0.9\textwidth]{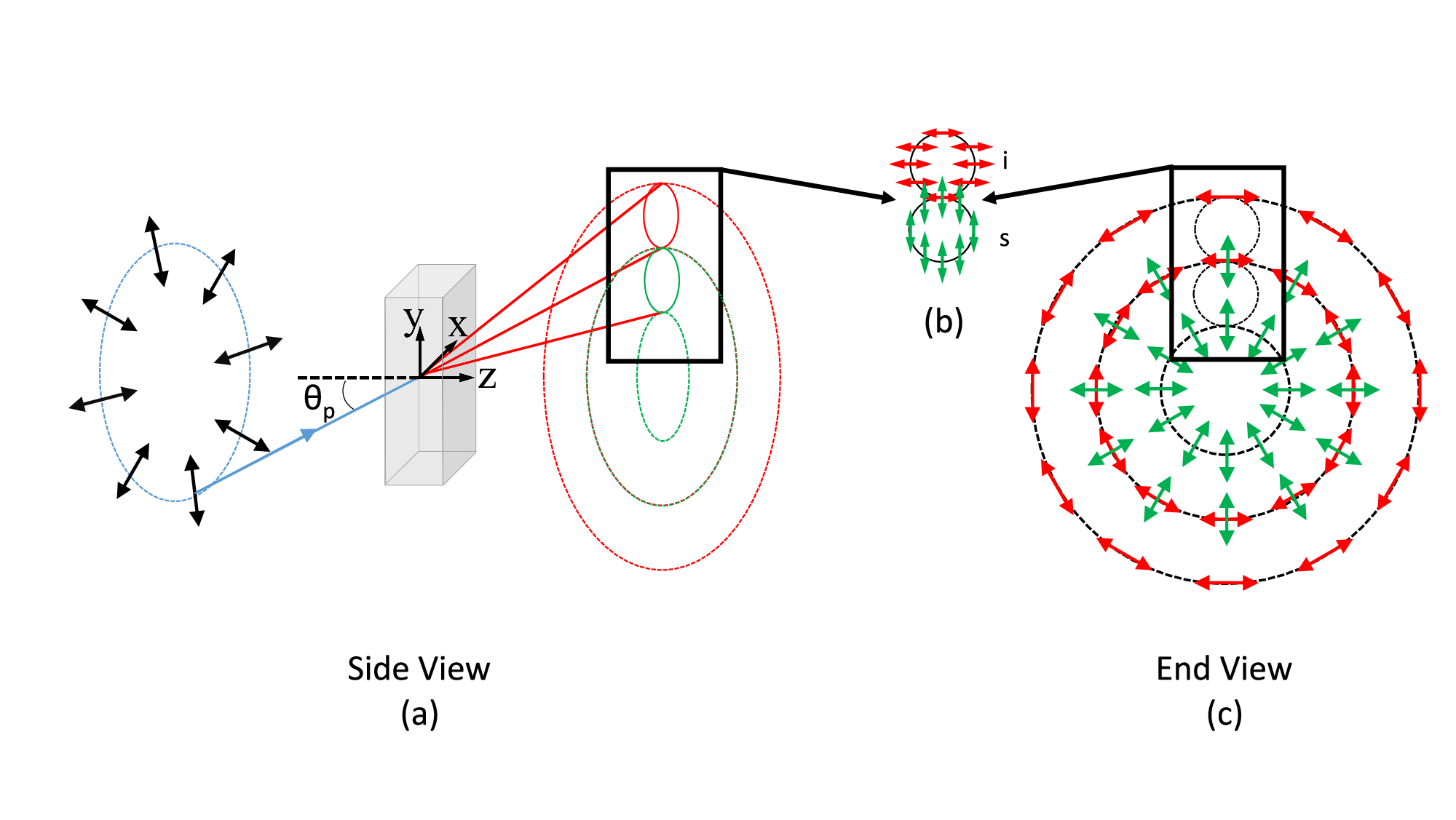}}
\caption{Spatial (a) and polarization (c) distributions for degenerate Type-II super-critical phasematching. The polarization distribution in (b) is the typical distibution for a single Gaussian pump beam. When this output is summed over all the Gaussian beams in the Bessel-Gauss pump distribution in Eq. (\ref{Eq:Bessel}), the resulting intensity and polarization distributions become those of (a) and (c). The signal photons (green) are generated with radial polarizations and the idler photons (red) are generated with azimuthal polarizations.}
\label{fig:ExpectedOutputTypeII}
\end{figure}

If we consider one such Gaussian pump beam, labelled by its central $\mathbf{k}$-vector $\mathbf{k}_p^0$ in Fig. \ref{fig:ExpectedOutputTypeII}(a), then the signal and idler photons produced by this Gaussian beam will be generated along the standard Type-II cones, shown in Fig.\ref{fig:ExpectedOutputTypeII}(b).  We can then take this standard Type-II SPDC output and apply the superposition principle to arrive at the expected output distribution for a Bessel-Gauss pump. In more detail, this output can be visualized by taking the two standard Type-II cones and rotating them around the crystal optical axis ($\mathbf{z}$-axis), as illustrated in Fig. \ref{fig:ExpectedOutputTypeII}(c). Rotation of the idler standard cone results in two larger and concentric cones, now centered on the $z$-axis. These are the supercones we observe in Fig. \ref{fig:TypeIIOutput}.  Similar supercones are produced by the rotation of the signal standard cone, one with the same diameter as the inner idler supercone and the other with a smaller diameter. Photon pairs may be produced in the regions between these supercones, but with lower probability, as can be seen in the profiles in Fig. \ref{fig:TypeIIOutput}. 

In this section, we have shown that the supercones seen in the simulation results have an intuitive geometric origin that can be understood by considering the standard output distributions for Type-II phasematched SPDC. Here, we only discussed as an example the case in which  $\theta_p$ is set to $41.8^\circ$. Different pump cone half-angles will change the diameter and intersection of the standard Type-II SPDC output circles. Using our intuitive geometric argument, one would expect more supercones to appear, two for each standard cone. 

Up to this point, we have only discussed the signal and idler marginal probability densities. In the next section, we examine the signal and idler joint probability density.

\section{Correlations}

\begin{figure}[!h]
\centering
\makebox[\textwidth][c]{\includegraphics[width=0.75\textwidth]{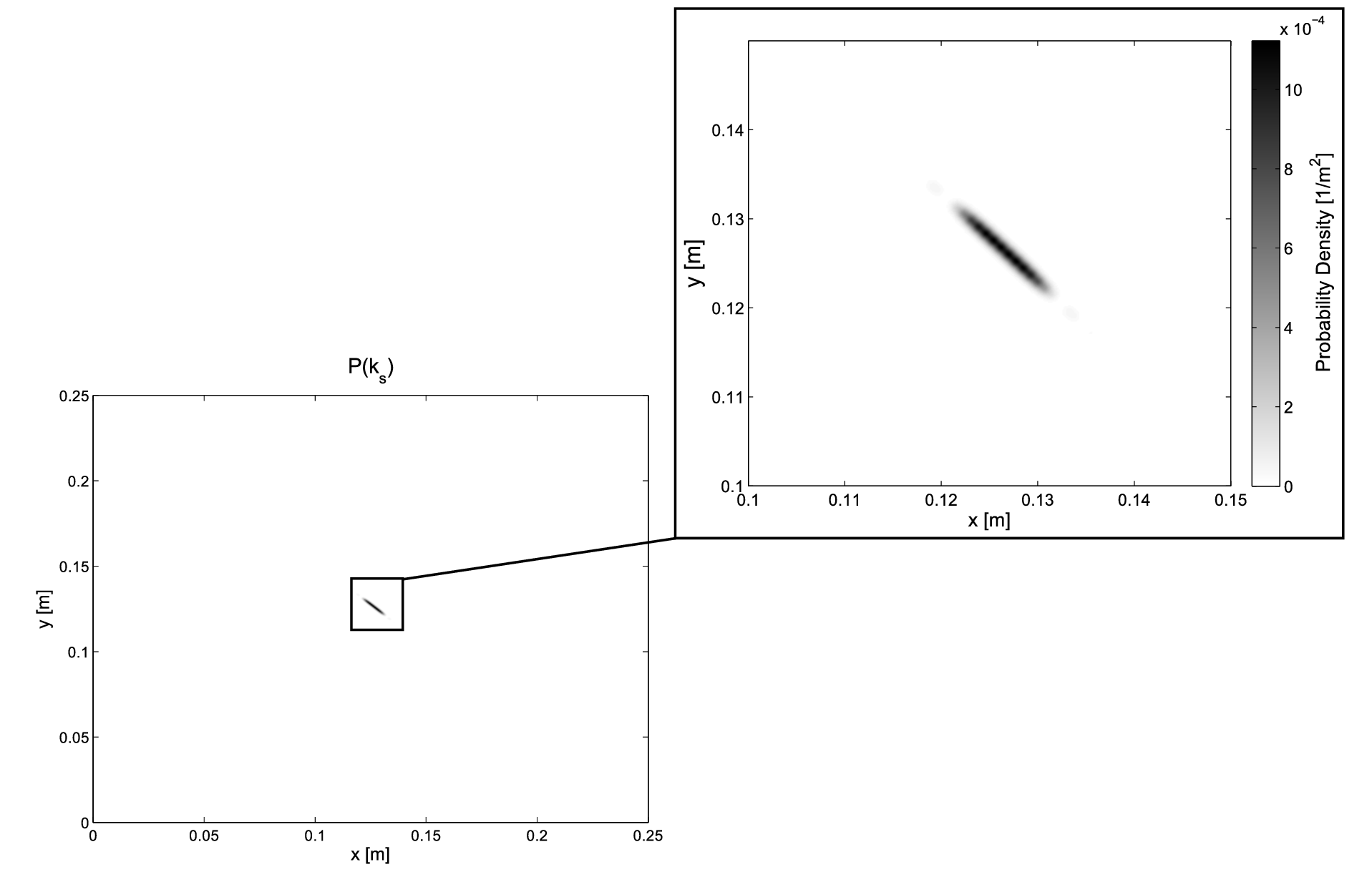}}
\caption{Signal photon probability density for Type-II degenerate collinear downconversion with a fixed idler emission direction. Idler angles are fixed at $\theta_i = 41.8^{\circ}$ and $\varphi_i = 45^{\circ}$.}
\label{SignalDistribution}
\end{figure}

One method to determine the correlations between the signal and idler photons is to calculate the full 4d signal and idler probability density $|\Phi_{BG}(\mathbf{k}_s, \mathbf{k}_i)|^2$. This would be achieved by calculating the phasematching amplitude in Eq. (\ref{Eq:PhiBG}) using 4d matrices of $\varphi_s$, $\theta_s$, $\varphi_i$, $\theta_i$ to give the phasematching amplitude for all possible combinations of these emission angles for the signal and idler. However, this calculation would be prohibitively large, so we instead simulate the probability density for the signal photon given the idler photon is found along $\varphi_i = 45^\circ$. We limit our discussion to the overlapping signal and idler supercone. That is, $\theta_{i}$ is chosen to be at the peak probability of this supercone. The results are shown in Fig. \ref{SignalDistribution}. 
 
The resulting signal conditional probability density has a FWHM width of $7.9$ mm in the $y'$ direction (azimuthal), which is 1.3\%  of the circumference of the signal supercone.  In the $x$' direction (radial), the conditional probability density has a width of $2.7$ mm, which is 39\% of the supercone FWHM thickness. Switching the role of signal and idler, these parameters are the same.  It is evident that signal and idler emission directions are strongly correlated: $\mathbf{k}_s$ eqals $\mathbf{k}_i$  to within the phasematching uncertainty set by Eq. (\ref{Eq:PhiG}). Both photons appear in the same spot in the overlapping supercone.

The strong correlations in the simulation imply that the full signal-idler two-photon wavefunction is highly entangled both in polarization and in spatial mode. In this situation, the position of, say, the idler photon reveals the position of the signal. In quantum mechanics, the presence of this "which-position" information requires absence of coherence between positions. Consequently, one cannot consider either photon, by itself, to be in a Bessel-Gauss quantum state, which would require coherence across the supercone. In particular, focusing the idler or signal photon supercone will result in a much broader distribution than what one would expect from Bessel-Gauss mode of this diameter. This is despite the fact that the marginal probability densities (Fig. \ref{fig:TypeIIOutput}) and polarizations (Fig. \ref{fig:ExpectedOutputTypeII}) of the signal and idler appear similar to those of radially and azimuthally polarized Bessel-Gauss modes, respectively.
 
On the other hand, applications in quantum information rely on entanglement and also on the symmetry of the overall two-photon state.  This is why we expect that the highly-entangled cylindrically symmetric two-photon states that this geometry produces to have many uses in this area.
 
\section{Feasibility of experimental realization}

The construction of a super-critically phasematched SPDC source will require consideration of refraction at the crystal face. While the axis of the pump cone is perpendicular to the crystal face, its opening angle, and hence its incidence angle, can be large. For the scenario discussed above, the opening half-angle of the pump cone is $\theta_p=41.8^\circ$ inside the crystal. This is larger than the angle of total internal reflection for BBO, which is $38^\circ$ at 405 nm. To circumvent this problem, we suggest two means of reducing the angle of the pump beam relative to the crystal face. 

\begin{figure}[h!t!p!b!]
\centering
\makebox[\textwidth][c]{\includegraphics[width=0.9\textwidth]{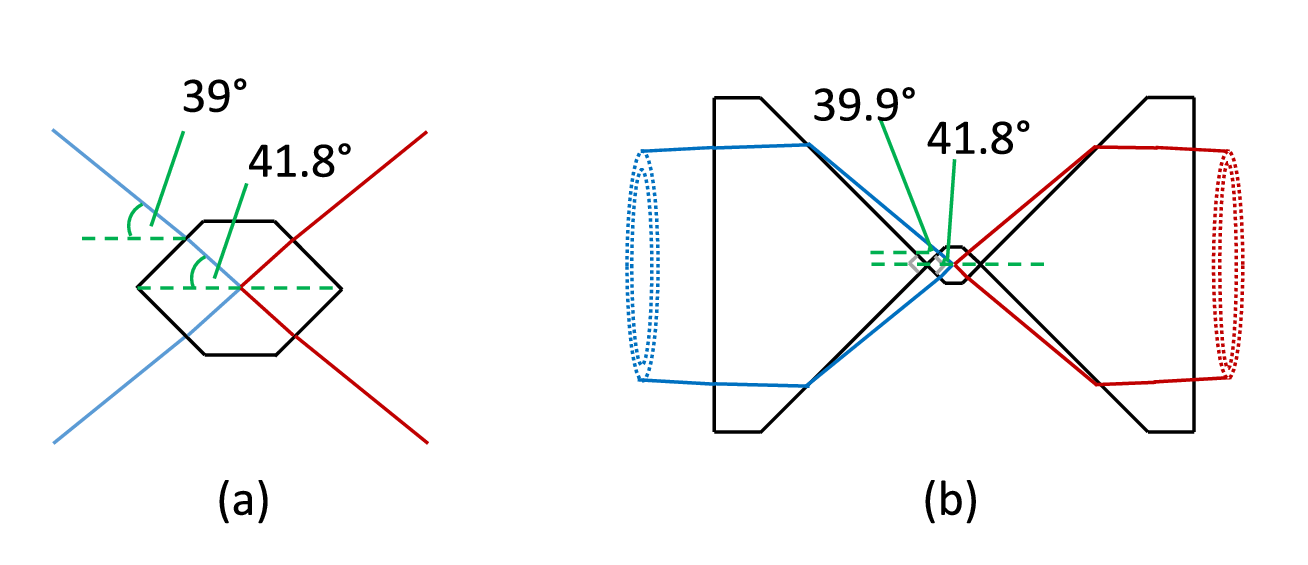}}
\caption{Proposed setup to reduce the pump opening angle in air. Angles are calculated for Type-II degenerate SPDC from 405 nm to 810 nm. The pump beam angle is reduced by two means: (a) a crystal cut as a double-sided $90^\circ$ axicon, and (b) two $90^\circ$ axicons placed tip-to-tip with the crystal at the entry and exit faces. The pump beam is focused into the crystal as it passes through the first axicon. The second axicon reduces the angle of the output beams as they exit the crystal.}
\label{fig:TypeII41.9Axicon}
\end{figure}

The first of these is to cut each face of the crystal as an axicon. That is, the crystal would be a double-sided cone with an apex full-angle of $90^\circ$. This reduces the pump beam angle to $5.1^\circ$ relative to the surface of the crystal, as shown in Fig. \ref{fig:TypeII41.9Axicon}. However, the pump angle is still quite large, $39.9^\circ$ relative to the $z$-axis, so another $90^\circ$ axicon is required to direct a diverging Bessel-Gauss pump beam (opening angle $3^\circ$) into the crystal. The pump beam is focused into the crystal when it passes through this axicon. An axicon is also placed at the exit face of the crystal to reduce the angle of the emitted cones of pump, signal, and idler photons.

\begin{figure}[h!t!p!b!]
\centering
\makebox[\textwidth][c]{\includegraphics[width=0.55\textwidth]{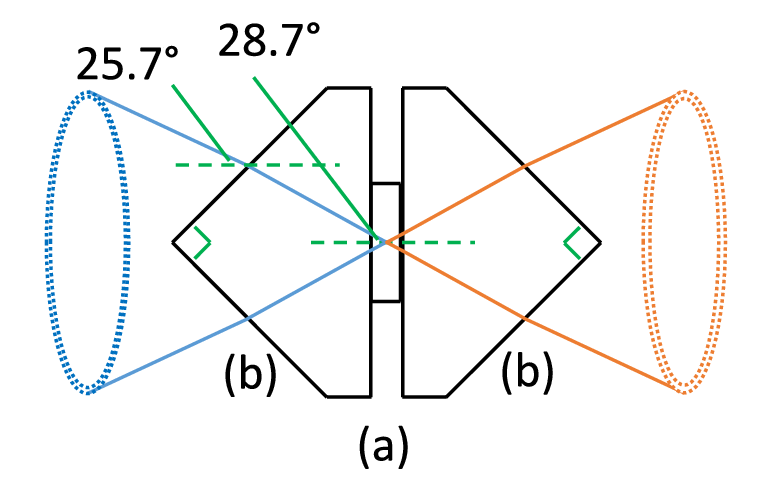}}
\caption{Proposed setup to reduce the pump opening angle in air for Type-II degenerate SPDC from 775 nm to 1550 nm: (a) a BBO crystal, placed between (b) two $90^\circ$ axicons, which are affixed back-to-back on either side of the crystal. The required opening angle of the pump in air is $25.7^\circ$; without the axicons at the crystal face this angle would be $51.8^\circ$.}
\label{fig:TypeII28.7Axicon}
\end{figure}

However, this method is not ideal because the cut of the crystal is not conventional and may be difficult to fabricate. Another method is to reduce the phasematching angle, $\theta_p$, by performing Type-II downconversion at a longer pump wavelength. To achieve the same standard Type-II geometry as discussed in Section 2 but for SPDC from 775 nm to 1550 nm, the phasematching angle is $\theta_p=28.7^\circ$. This is less than the angle of total internal reflection for BBO, which avoids the need for an unusual crystal shape. However, due to refraction at the crystal face the pump beam must be focused into the crystal at an angle of $51.8^\circ$ to meet phasematching conditions inside the crystal. This focusing angle is difficult to achieve, but can be reduced by affixing two $90^\circ$ axicons back-to-back on either side of the crystal, as shown in Fig. \ref{fig:TypeII28.7Axicon}. The index of glass is similar to that of BBO so the pump beam does not refract significantly when entering the crystal through the axicon. Thus, a pump beam that is focused at $25.7^\circ$ relative to the horizontal is sufficient to produce an angle of $28.7^\circ$ inside the crystal. The axicon affixed to the exit face of the crystal similarly reduces the exit angles of the pump, signal, and idler supercones.

The use of periodically poled KTP was also considered as a means of reducing the entry angle into the crystal for Type-II phasematching. This would require that the crystal be poled with the optical axis parallel to the beam propagation axis; however, current poling methods only allow for the optical axis to be perpendicular to the direction of propagation. KTP is also biaxial, which means that the crystal principal indices of refraction in the $x$- and $y$- directions are not exactly equal at a given wavelength. As a result, the pump azimuthal angle $\varphi_p$ would affect the phasematching conditions and the cylindrical symmetry would be lost.

An additional experimental consideration is the polarization of the pump beam. Due to the difficulty in producing a radially polarized pump beam, it may be advantageous to instead produce a circularly polarized Bessel-Gauss beam with an OAM value of $\pm 1$, the sign being opposite to the handedness of the polarization. This beam has equal azimuthal and radial polarization components. Only the radial component will contribute to the production of photon pairs because this is the extraordinary component of the pump beam. As a result, the contributing intensity will be half that of the total intensity. 

\section{Conclusion}

We have described a novel, cylindrically symmetric geometry for spontaneous parametric downconversion with the aim of producing radially and azimuthally polarized photon pairs. The pump beam is a radially (or alternatively, circularly) polarized Bessel-Gauss beam, which can be thought of as a distribution of Gaussian beams with central $\mathbf{k}$-vectors forming the surface of a cone. The pump beam is focused into a nonlinear crystal such that the cone axis is along the optical axis of the crystal.

We numerically simulated the output distributions of the signal and idler photons. The photons emerge along cylindrically symmetric (about the optical crystal axis) distributions which we call supercones. The signal photons will be emitted with radial polarizations and the idler photons will have azimuthal polarizations. The photons exhibit strong correlations such that the signal and idler appear in the same location in the central supercone. Consequently, the associated two-photon wavefunction is strongly entangled both in polarization and in spatial mode.

While not the focus of this paper, this beam geometry could potentially improve the efficiency of the reverse process, second harmonic generation (SHG), in which two photons combine to produce one photon at a higher frequency. As this requires the presence of two photons, the efficiency of SHG depends quadratically on the intensity of the beam. Focusing the beam more tightly to increase this intensity causes the beam to diffract more quickly, so the crystal length is limited by the Rayleigh range of the beam \cite{Boyd1968}. Bessel beams may provide improved conversion efficiencies over Gaussian beams due to their non-diffracting properties, in that they maintain a high peak intensity over larger distances \cite{shinozaki1997comparison}. Attempts have been made to experimentally test this idea. In particular, truncated Bessel beams were used to pump Type-I phasematching in lithium triborate \cite{physRevA.60.2438}. In that paper, measured conversion efficiencies were slightly lower than those obtained for a Gaussian pump. However, that geometry consisted of an input beam traveling perpendicular to the crystal axis in \textit{non-critical} phasematching, and showed only longitudinal or transverse phasematching, but not both concurrently. As this is significantly different from our \textit{super-critical} phasematching, it is worth again considering whether Bessel beams can enhance SHG efficiency.

\section*{Funding} 

This research was undertaken, in part, thanks to funding from the Canada Research Chairs, NSERC Discovery, Canada Foundation for Innovation, and the Canada Excellence Research Chairs program.

\section*{Appendix A1}

\begin{figure}[h!t!p!b!]
\makebox[\textwidth][c]{\includegraphics[width=1\textwidth]{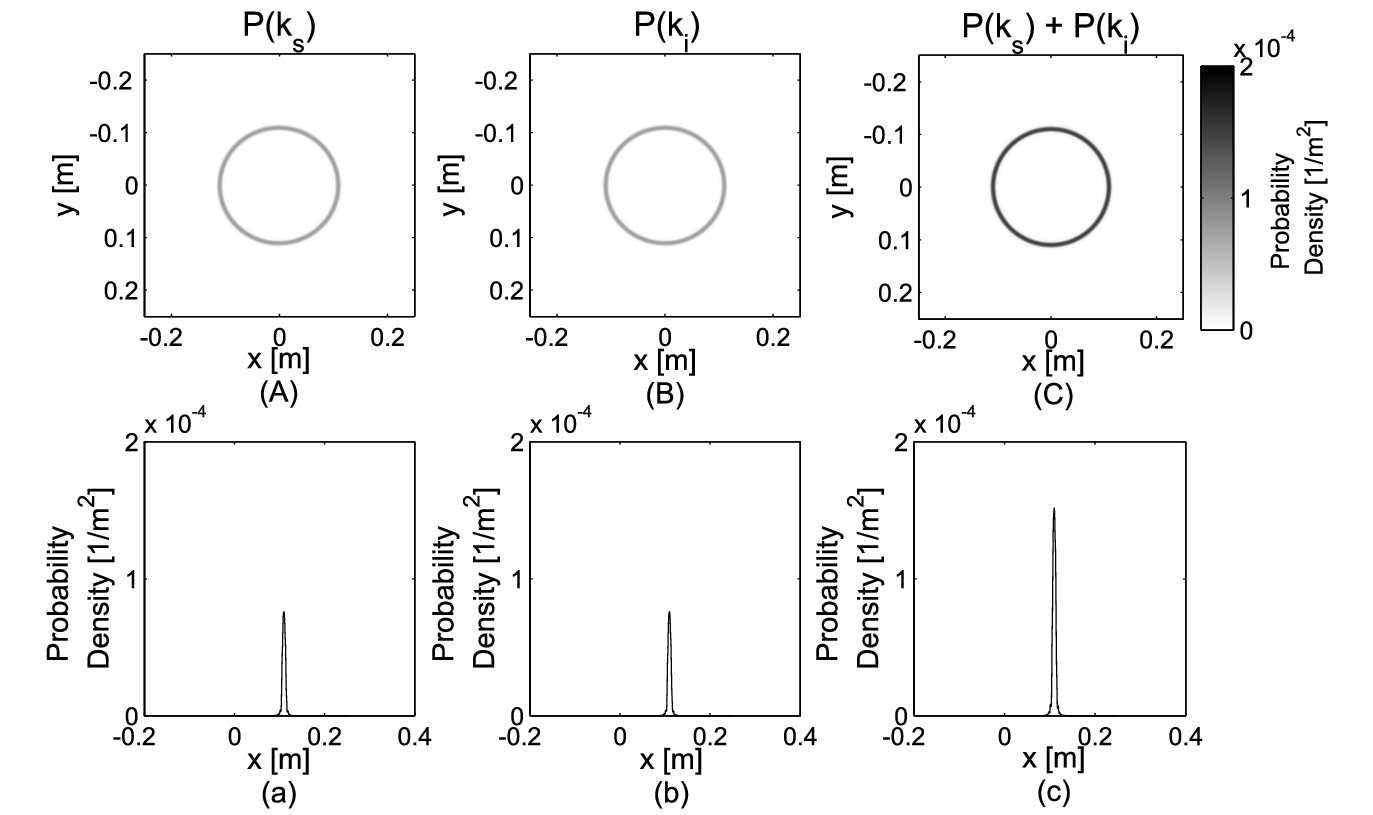}}
\caption{Probability densities for Type-I degenerate downconversion in a $500\ \mu$m -thick BBO crystal for a pump beam wavelength of 405 nm, and signal and idler wavelength of 810 nm. Since for degenerate phasematching $|\mathbf{k}_{s(i)}|$ is fixed, $\theta_{s(i)}$ and $\phi_{s(i)}$ completely determine the signal and idler wavevectors. In turn, the position of the photons after the crystal determines these angles, so these plots are given in the crystal frame (x,y) at z=20cm after the crystal exit face. Plots (A), (B), and (C) give (A) the probability density for the signal photon, $P(\mathbf{k}_s)$; (B) the probability density for the idler photon, $P(\mathbf{k}_i)$; and (C) $P(\mathbf{k}_s)$ + $P(\mathbf{k}_i)$, which is proportional to the photon flux. Plots (a), (b), and (c) are produced from plots (A), (B), (C), respectively, by plotting along $y = 0$.}
\label{fig:TypeIOutput}
\end{figure}

We consider here the case of Type-I degenerate phasematching. Fig. \ref{fig:TypeIOutput} shows the simulated probability densities $|\Phi_{BG}|^2$ for the signal and idler, where $\Phi_{BG}$ is defined as in Eq. (\ref{Eq:PhiBG}). The pump opening half-angle  $\theta_p$  was set to $28.8^{\circ}$ inside the crystal. In the standard Type-I geometry, this is the phasematching angle that would produce signal and idler photons collinear with the pump beam. In contrast, for a Bessel-Gauss pump the resulting SPDC output distribution is a single supercone that is at all points collinear with the pump cone. 

\begin{figure}[h!t!p!b!]
\centering
\makebox[\textwidth][c]{\includegraphics[width=0.9\textwidth]{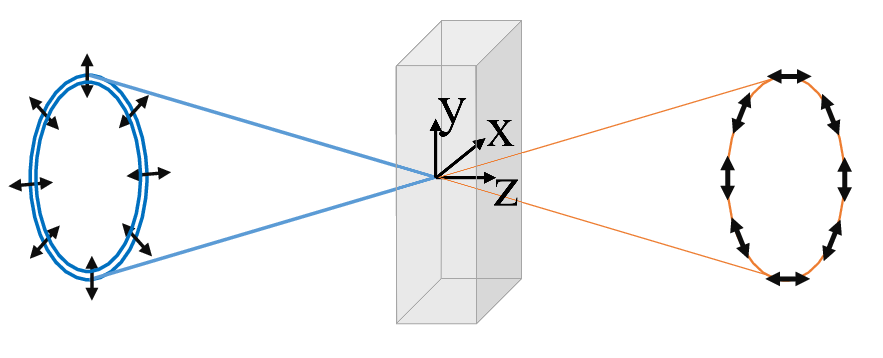}}
\caption{Expected output for Type-I collinear downconversion with a conical pump beam centered on the optic axis (\textit{z}). The signal and idler photons are produced in a single supercone centered on the optic axis.}
\label{fig:ExpectedOutputTypeI}
\end{figure}

This output distribution can be understood by superposing the typical Type-I SPDC outputs for each Gaussian beam in the pump distribution in Eq. (\ref{Eq:PhiBG}). Each of these Gaussian beams produces collinear signal and idler photons. Consequently, only one supercone is produced containing both signal and idler photons.  For Type-I phasematching, the signal and idler are emitted with ordinary polarizations. When these outputs are superposed for all Gaussian beams in the pump distribution, the resulting polarization distribution is azimuthal. This polarization distribution is shown in Fig. \ref{fig:ExpectedOutputTypeI}.

The Type-I phasematching angle is below the angle of total internal reflection for BBO at 405 nm; however, there is still significant refraction at the crystal face so a reduction in the free-space opening angle of the pump beam is required. The pump beam can be focused into the crystal using a similar setup as in Fig. \ref{fig:TypeII28.7Axicon}.

\section*{Appendix A2}

The output distributions shown in Figs. \ref{fig:TypeIIOutput} and \ref{fig:TypeIOutput} were calculated in Matlab and took 8 days to run on a core i5 processor. For most of the work below we follow the work of Ref. \cite{boeuf2000calculating}. We start with a 750 by 750 grid of (x,y) positions, with a range large enough to accommodate all pump, signal, and idler \textbf{k}-vectors. That is, from $-0.25$ m to $0.25$ m. We then calculate the downconversion amplitudes for each Gaussian beam in the pump distribution in Eq. (\ref{Eq:PhiBG}). We then sum these amplitudes, squaring the result to give the total probability density. The procedure for a single Gaussian pump beam is given below, comprising Eqs. (\ref{Eq:Theta_s,i}) to (\ref{Eq:Amplitude}). Starting with the grid of (x,y), angles $\theta_{s(i)}$ and $\varphi_{s(i)}$ were calculated for the signal and idler photons as follows:

\begin{equation}
\theta_s = \tan^{-1}\left({\frac{\sqrt{{x}^2 + {y}^2}}{{z}}}\right), \\
\theta_i =-\theta_s,
\label{Eq:Theta_s,i}
\end{equation}

\begin{equation}
\varphi_s = \tan^{-1}\left(\frac{{x}}{{y}}\right), \\
\varphi_i = \varphi_s + \pi,
\label{Eq:Phi_s,i}
\end{equation}

\noindent
where $\theta_{s,i}$ and $\varphi_{s,i}$ are the angles shown in Fig. \ref{fig:SPDCGeo}, and are defined relative to the optic axis, and z is set to be 0.2 m. 

The components of the propagation directions $\mathbf{s}_{p/s/i}$ for the pump, signal, and idler, were calculated from $\theta$ and $\varphi$ as follows:

\begin{equation}
s_{\alpha, x} = \sin(\theta_{\alpha})\cos(\varphi_{\alpha}),\ \  \\
s_{\alpha, y} = \sin{(\theta_{\alpha})}\sin{(\varphi_{\alpha})},\ \  \\
s_{\alpha, z} = \cos{(\theta_{\alpha} )},
\end{equation}

\noindent
where $\alpha = p,s,i$.  Additionally, the crystal principal indices of refraction $n_x$, $n_y$, and $n_z$ were calculated for the pump, signal, and idler photons. The Sellmeier equations for BBO were obtained from \cite{boeuf2000calculating}. The extraordinary and ordinary indices of refraction are then:

\begin{equation}
N_{\alpha,e} = {\left(\frac{2}{B + {(B^{2} - 4C)}^{1/2}}\right)}^{1/2}
\end{equation}
\begin{equation}
N_{\alpha,o} = {\left(\frac{2}{B - {(B^{2} - 4C)}^{1/2}}\right)}^{1/2},
\end{equation}

\noindent
where $\alpha = p, s, i$ and

\begin{equation}
B = \frac{s_{\alpha, x}^2}{n_{\alpha, y}^2 + n_{\alpha, z}^2} + \frac{s_{\alpha, y}^2}{n_{\alpha, x}^2 + n_{\alpha,y}^2} + \frac{s_{\alpha,z}^{2}}{n_{\alpha,x}^2 + n_{\alpha,y}^2},
\end{equation}
\begin{equation}
C = \frac{s_{\alpha,x}^2}{n_{\alpha,y}^{2}n_{\alpha,z}^{2}} + \frac{s_{\alpha,y}^2}{n_{\alpha,x}^{2}n_{\alpha,z}^{2}} + \frac{s_{\alpha,z}^2}{n_{\alpha,x}^{2}n_{\alpha,y}^{2}}.
\end{equation}

\noindent
The extraordinary or ordinary indices were assigned to the signal and idler photons depending on which type of phasematching, Type-I or Type-II, was chosen. We can then determine the \textbf{k}-vector components for the pump, signal, and idler as follows:

\begin{equation}
k_{\alpha, j} = 2\pi\frac{N_{\alpha}s_{\alpha,j}}{\lambda_{\alpha}},
\label{Eq:Kj}
\end{equation}

\noindent
in which $\alpha = p,s,i$ and ${j = {x,y,z}}$. The direction vector components ${[s_x, s_y, s_z]}$ are defined with respect to the optic axis, so these \textbf{k}-vectors are given in the crystal frame. The \textbf{k}-vector mismatches in the crystal frame of reference are then given by Eq. (\ref{Eq:delK}):

\begin{equation}
\Delta k_{j} = k_{s,j} + k_{i,j} - k_{p,j},
\label{Eq:delK}
\end{equation}

\noindent
where j = x,y,z. So far, we have worked in the crystal frame because it is convenient to have a consistent frame of reference for plotting. However, due to the non-paraxial nature of the Bessel-Gauss pump beam, we have chosen to calculate the phasematching amplitudes in the local pump frame ($\mathbf{x}'$, $\mathbf{y}'$, $\mathbf{z}'$) for each paraxial Gaussian beam in the pump distribution (see Fig. \ref{fig:SPDCGeo}). To do so, we first convert the \textbf{k}-vector mismatches from the crystal frame ($\mathbf{x}$, $\mathbf{y}$, $\mathbf{z}$) to the pump frame ($\mathbf{x}'$, $\mathbf{y}'$, $\mathbf{z}'$). The transformations between these coordinate systems are given by:

\begin{equation}
\left[\begin{array}{c}
x \\
y \\
z \\
\end{array}\right] = \left[\begin{array}{ccc}
\cos(\theta_p)\cos(\varphi_p) & \cos(\theta_p)\sin(\varphi_p) & -\sin(\varphi_p) \\
-\sin(\varphi_p) & \cos(\varphi_p) & 0 \\
\sin(\theta_p)\cos(\varphi_p) & \sin(\theta_p)\sin(\varphi_p) & \cos(\theta_p) \\
\end{array}\right] \left[\begin{array}{c}
x' \\
y' \\
z' \\
\end{array}\right].
\label{Eq:crystaltopumpframe}
\end{equation}

The transformation from the $\mathbf{k}$-vector mismatches in the crystal frame to the $\mathbf{k}$-vector mismatches in the pump frame is then given by:

\begin{equation}
\Delta k_{x'} = \Delta k_{x}\cos(\theta_p)\cos(\varphi_p) + \Delta k_{y}\cos(\theta_p)\sin(\varphi_p) - \Delta k_{z}\sin(\varphi_p),
\end{equation}
\begin{equation}
\Delta k_{y'}= -\Delta k_{x}\sin(\varphi_p) + \Delta k_{y}\cos(\varphi_p),
\end{equation}
\begin{equation}
\Delta k_{z'} = \Delta k_{x}\sin(\theta_p)\cos(\varphi_p) + \Delta k_{y}\sin(\theta_p)\sin(\varphi_p) + \Delta k_{z}\cos(\theta_p).
\label{Eq: Delta k_pump}
\end{equation}

These \textbf{k}-vector mismatches are substituted into Eq. \ref{Eq:Amplitude}:

\begin{equation}
\Phi_G(\mathbf{k}_p^0, \mathbf{k}_s, \mathbf{k}_i) = \exp\left({\frac{-w_{p}^{2}(\Delta k_{x'}^2 + \Delta k_{y'}^2)}{4}}\right)\textrm{sinc}\left(\frac{L_{\mathrm{optic}}\Delta k_{z'}}{2}\right)\exp\left({\frac{iL_{\mathrm{optic}}\Delta k_{z'}}{2}}\right),
\label{Eq:Amplitude}
\end{equation}

\noindent
in which $\Delta k_{x'}$, $\Delta k_{y'}$, and $\Delta k_{z'}$ are the \textbf{k}-vector mismatches in the \textit{x'}, \textit{y'}, and \textit{z'} directions. Here, $L_{\mathrm{optic}}$ is the length of the optical path through the crystal. That is, $L_{\mathrm{optic}}=L_{\mathrm{crystal}}/\cos(\theta_p)$, where $L_{\mathrm{crystal}}$ is the crystal length as measured along its normal. Eq. (\ref{Eq:Amplitude}) gives the complex amplitude $\Phi_G(\mathbf{k}_p^0, \mathbf{k}_s,\mathbf{k}_i)$ to generate a signal and idler photon at $\mathbf{k}_s$ and $\mathbf{k}_i$, respectively, from a Gaussian pump beam centered on $\mathbf{k}_p^0$. The first exponential in Eq. (\ref{Eq:Amplitude}) corresponds to phasematching transverse to the central pump \textbf{k}-vector. Similarly, the sinc term corresponds to phasematching in the longitudinal direction. The last exponential determines the phase. The corresponding probability is given by $|\Phi_G(\mathbf{k}_p^0, \mathbf{k}_s, \mathbf{k}_i)|^2$.

The plots shown in  Fig.~\ref{fig:TypeIIOutput} and  Fig.~\ref{fig:TypeIOutput} were produced from the above calculations, with the distinctions between these plots given as follows:

Figures \ref{fig:TypeIIOutput}(A) and \ref{fig:TypeIOutput}(A) give the marginal probability densities for the signal for Type-II and Type-I downconversion, respectively. The amplitude to produce a signal at $\mathbf{k}_s$ was first calculated for a single central Gaussian pump $\mathbf{k}$-vector $\mathbf{k}_p^0$. For each value of $\mathbf{k}_s$, $\Phi_G$ was determined for all idler $\mathbf{k}$-vectors in the grid. This calculation was then performed for all central $\mathbf{k}$-vectors in the pump distribution, and the outputs were then summed to give the amplitude to produce a photon at our value of $\mathbf{k}_s$ given a Bessel-Gauss pump. We have now determined the amplitude $\Phi_{BG}(\mathbf{k}_s, \mathbf{k}_i)$, given by:

\begin{equation}
\Phi_{BG}(\mathbf{k}_s, \mathbf{k}_i) \propto \sum_{\varphi_p}\Phi_G(\mathbf{k}_p^0(\varphi_p), \mathbf{k}_s, \mathbf{k}_i),
\label{Eq:PhiBG_ks}
\end{equation}

\noindent
where we have approximated the integral over $\mathbf{k}_p^0$ in Eq. (\ref{Eq:PhiBG}) with a sum over 750 values of $\mathbf{k}_p^0$ with $\varphi_p$ ranging from 0 to $2\pi$. The marginal probability density for the signal photon is then obtained by taking the absolute square of this amplitude:

\begin{equation}
P(\mathbf{k}_s) = \frac{1}{N}\sum_{\mathbf{k}_{i,x}, \mathbf{k}_{i,y}}|\Phi_{BG}(\mathbf{k}_s, \mathbf{k}_i)|^2\Delta \mathbf{k}_{i,x}\Delta \mathbf{k}_{i,y},
\label{Eq:PS}
\end{equation}

\noindent
in which N is a normalization factor such that $\sum_{\mathbf{k}_s} P(\mathbf{k}_s)\Delta\mathbf{k}_s = 1$. An identical procedure was used to determine the marginal probability densities $P(\mathbf{k}_i)$ for the idler photon, shown in Figs. \ref{fig:TypeIIOutput}(B) and \ref{fig:TypeIOutput}(B).

\end{document}